\documentclass[aps,prd,twocolumn,a4paper,showpacs,superscriptaddress]{revtex4-1}
\usepackage{pgf}
\usepackage[T1]{fontenc}
\usepackage[latin1]{inputenc}
\usepackage{graphicx,amsfonts}
\usepackage{amsmath}
\newcommand{\be}{\begin{equation}}
\newcommand{\ee}{\end{equation}}
\newcommand{\bea}{\begin{eqnarray}}
\newcommand{\eea}{\end{eqnarray}}
\newcommand{\bml}{\begin{subequations}}
\newcommand{\eml}{\end{subequations}}

\newcommand{\bbm}{\begin{bmatrix}}
\newcommand{\ebm}{\end{bmatrix}}
\newcommand{\sech}{\mathrm{~sech}}

\newcommand{\arctanh}{\mathrm{arctanh}}

\begin{document}
\title{From Supersymmetric Quantum Mechanics to Scalar Field Theories}
\author{D. Bazeia}
\email{bazeia@fisica.ufpb.br}
\affiliation{Departamento de F\'\i sica, Universidade Federal
da Para\'\i ba, 58051-970, Jo\~ao Pessoa, PB, Brazil}
\author{F. S. Bemfica}
\email{fabio.bemfica@ect.ufrn.br} \affiliation{Escola de Ci\^encias
e Tecnologia, Universidade Federal do Rio Grande do Norte, 59072-970, Natal, RN, Brazil}
\date{\today}

\begin{abstract}
In this work we address the reconstruction problem, investigating the construction of field
theories from supersymmetric quantum mechanics. The procedure is reviewed, starting from reflectionless
potentials that admit one and two bound states. We show that, although the field theory reconstructed
from the potential that supports a single bound state is unique, it may break unicity in the case of two bound states.
We illustrate this with an example, which leads us with two distinct field theories.
\end{abstract}

\maketitle

\section{Introduction}

Reconstruction of field theories from reflectionless symmetric quantum mechanical
potentials is an old issue~\cite{Jackiw:1977yn,Christ:1975wt,Boya:1989db,Casahorran:1990ck,FloresHidalgo:2002ma,Junker:1996cf,Cooper:1994eh,Kwong:1985ti,
Bordag:2002pm,Vachaspati:2003vk}.
An interesting fact involving reflectionless potentials is that it can be
constructed univocally once the bound state spectrum is known, in a procedure that
is sometimes called the spectral method~\cite{Junker:1996cf,Cooper:1994eh}. 

In the case of field theories that support defect structures, the study of stability
is directly connected to supersymmetric quantum mechanics~\cite{Jackiw:1977yn}. The interesting
question that arises in such situation is whether there is a field
theory model associated with each supersymmetric quantum
mechanical potential. Moreover, such question can be enriched by
asking if, given the spectrum of a quantum mechanical
potential, a field theory model is reconstructed. The answer is
not new and has been given in~\cite{Bordag:2002pm,Vachaspati:2003vk} for a couple of
reflectionless quantum mechanical examples. See also
Refs.~\cite{Christ:1975wt,Boya:1989db,Casahorran:1990ck,FloresHidalgo:2002ma} for other details.
An open questions that remains to be studied is whether this
reconstruction is unique. The
answer to this question is considered in this work.

Throughout the current investigation, we shall be dealing with a real scalar field
$\phi=\phi(x,t)$ in the $(1,1)$-dimensional spacetime, described by the action
\be\label{1}
S[\phi]=\int\,dt\,dx
\Bigl[\frac{1}{2}\partial_\mu\phi\partial^\mu\phi-V(\phi)\Bigr]\,.
\ee
Here, $\mu=0,1$ stand for time and space coordinates,
respectively, with $x^0=t$ and $x^1=x$, while $V$ is some function of the field $\phi$.
In many cases, it can be constructed via another function $W=W(\phi)$, in the form
\be\label{2}
V(\phi)=\frac12 W^2_\phi\,,
\ee
where $W_\phi\equiv dW/d\phi$. The field $\phi$, the
space $x$, and the time $t$ are redefined here in such a way that
they are all dimensionless, so the work is written using
dimensionless quantities. The equation of motion that appears from the
action (\ref{1}) is
\be
\label{3}
\ddot{\phi}-\phi^{\prime\prime}+V_\phi=0\,,
\ee
where the dot is a time derivative while the prime means $x$ derivative.
Static solutions of (\ref{3}) obey
\be
\label{4}
\phi^{\prime\prime}=V_\phi\,.
\ee
When the potential is written in terms of $W$, as in Eq.~\eqref{2}, we are interested in the solutions of the first-order differential equations
\be
\label{5}
\phi^\prime=\pm W_\phi\,,
\ee
known as Bogomol'nyi, Prasad and Somerfield (BPS) equations\cite{Bogomolnyi,Prasad}. The signs in the above equations
are used to distinguish between kinks and antikinks.

Since supersymmetric quantum mechanical models arise when one addresses the
stability of the static solution, we then consider
small fluctuation of the time-dependent solution $\phi(x,t)$
around the static solution $\phi(x)$; namely, we take $\eta(x,t)\approx
\phi(x,t)-\phi(x)$. In this case, $\eta(x,t)$ obeys the partial
differential equation
\be\label{6}
\ddot\eta-\eta^{\prime\prime}+U(x)\eta=0,
\ee
where
\be
\label{7}
U(x)=V_{\phi\phi}=W_{\phi\phi}^2+W_{\phi\phi\phi}W_\phi,
\ee
is the stability potential. Since $U(x)$ only depends on $x$, because we are considering fluctuation around a static solution, we can separate variables and perform the mode expansion
\be
\label{8}
\eta(x,t)=\sum_{n=0}^\infty \eta_n(x)\cos(\omega_n t)\,.
\ee
In this case the resulting quantum mechanical problem can be cast into the form
\be
\label{9}
\Bigl[-\frac{d^2}{dx^2} +U(x)\Bigr]\eta_n(x)=\omega^2_n\eta_n(x)\,.
\ee
Now, stability demands $\omega_n^2\ge0$, which is guaranteed since
the Hamiltonian
\be
\label{10}
H=-\frac{d^2}{dx^2}+U(x)=\left(-\frac{d}{dx}+f\right)\left(\frac{d}{dx}+f\right)=S^\dagger S\,,
\ee
is factorized through the operator
\be
S=\frac{d}{dx}+f,
\ee
with $U(x)=f^2-f^\prime$. One may recognize from \eqref{7} that $f= \mp W_{\phi\phi}$. The sign $\mp$ for $f$
depends on the sign assigned to the first-order Eq.~\eqref{5}, and to $W$ itself. For completeness, the
supersymmetric partner of $H$ is
\be
H_{ss}=SS^\dagger=-\frac{d^2}{dx^2}+U_{ss}(x),
\ee
with $U_{ss}(x)=f^2+f^\prime$.

The field theory in \eqref{1} is translationally invariant, and this ensures
the existence of the translation or zero mode $\eta_t(x)$ in the
spectrum of $H$, with corresponding energy eigenvalue
$\omega_t^2=0$. One uses Eqs.~\eqref{9} and \eqref{10} to get $S\eta_t(x)=0$, which gives
\be
\label{11}
\eta_t(x)=N\exp\left(-\int f\, dx\right)=\frac{N}{w(x)}\,,
\ee
where $N$ is a normalization constant for $\eta_t(x)$. For future
purposes we have introduced the function $w(x)$, which must be
nonlimited as $x\to\pm\infty$; it is related to $f$ by the
definition
\be
\label{12}
f(x)=\frac{w^\prime}{w}\,.
\ee
A condition on $f$ for the existence of a zero
mode demands that $f_-=f(x\to-\infty)<0$ \cite{Cooper:1994eh}.

This paper is organized as follows. In Sec.~\ref{reconstruction}, we review and summarize
the reconstruction of a field theory with topological structures emerging from a
quantum mechanical problem as well as the process of
obtaining a quantum mechanical potential from a discrete energy
spectrum. We move on to investigate in Sec.~\ref{section3} the problem of
uniqueness of the reconstruction procedure, addressing the two
known problems that consider reflectionless potentials with one and
two bound states. Nonuniqueness of the reconstruction procedure is shown in the case with two bound states. 
Finally, in Sec.~\ref{conclusions} we add some comments and conclusions.

\section{Reconstruction scheme}
\label{reconstruction}

We follow the procedure of potential reconstruction given in
\cite{Kwong:1985ti}. Once the quantum mechanical potential is
obtained, the steps suggested in \cite{Bordag:2002pm,Vachaspati:2003vk} are used in
order to reconstruct the field theory model. To begin with,
let us sketch the reconstruction of the field potential $V(\phi)$
once one knows $U(x)$. If the $x$ derivative of (\ref{4}) is
performed, one may recognize that $\phi^\prime\propto \eta_t(x)$.
Then, given $U(x)$, Eq. (\ref{9}) for $\omega^2_t=0$ together with
(\ref{11}) enable us to write
\be
\phi^\prime=\pm\frac{\eta_t(x)}{N}=\pm\frac{1}{w(x)}\,,
\ee
which lead us to
\be
\label{2-1}
\phi=\pm\int \frac{dx}{w(x)}-c\,.
\ee
The $\pm$ sign in the above equation is to take into account the
two possible BPS solutions (\ref{5}), while $c$ is just an
integration constant. Now, if $x(\phi)$, the inverse of the
function $\phi(x)$ in \eqref{2-1}, can be obtained analytically,
it is straightforward to arrive at the potential
\be
\label{2-2}
V(\phi)=\left.\frac{\eta_t(x)^2}{2N^2}\right|_{x=x(\phi)}
=\left.\frac{1}{2w(x)^2}\right|_{x=x(\phi)}\,.
\ee

Here a digression is needed. Although it seems that the constant
of integration in (\ref{2-1}) can be absorbed by a simple
redefinition of the field $\phi$, corresponding to an equivalent
field theory model, one should take into account that, when the
model being reconstructed has more than one topological sector,
the constant $c$ may take more than one value, leading to other possible static solutions.
This is an important point, and it was raised after our previous
work on the subject \cite{Bazeia:2016xfc}. It precludes the elimination of $c$ from the
problem by performing a unique redefinition of $\phi$, which, depending on each case,
may give inequivalent actions (\ref{1}), that is, different scalar field theories. Now, extra
information is needed for the univocal reconstruction of the field theory, namely,
the existence of other topological sectors. We note that one possible value of $c$ is $c=0$; however, since
the solutions in \eqref{2-1} must identify its own sector, one may also take, for instance,
\be\label{check}
c=\pm\lim_{x\to\infty} \int \frac{dx}{w(x)},
\ee
and check if it brings new possibilities. Below, we illustrate this point with two distinct investigations.

Let us now summarize the reconstruction of $U(x)$ from its
scattering data for the case of a positive energy
spectrum \cite{Kwong:1985ti}. We first recall that this is always the case because we are reconstructing
field theory models based on the existence of non-negative potentials of the form \eqref{2}, and this implies 
that all the kinklike solutions are BPS states that solve first-order equations \cite{Bazeia:2001te}. Now, suppose we know the $N$ bound states
of $U(x)$ to be $\omega_1>\omega_2>\cdots>\omega_N=0$, the last
one being the smallest one and corresponding to the zero mode,
since the reconstructed field theory must be translationally
invariant. Also, let $U_n(x)$ be a potential that contains the
bound states $\omega_1>\omega_2\cdots>\omega_n$ for
$n=1,\cdots,N$. If $n\ne N$, then $\omega_n\ne 0$. However,
$U_n-\omega_n^2$ corresponds to a potential which contains
a zero mode. From supersymmetric quantum
mechanics~\cite{Cooper:1994eh}, the potential $U_n-\omega_n^2$ can
be obtained from $f_n$ such that
$U_n-\omega_n^2=f_n^2-f^\prime$, while its superpartner potential
$U_{n-1}(x)-\omega_n^2=f_n^2+f_n^\prime$ will have the same
spectrum except for the zero mode. In other words,
$U_{n-1}(x)$ will have the bound state energies
$\omega_1>\omega_2>\cdots>\omega_{n-1}$. This enables one to get
the recurrence equations
\bml
\label{2-3}
\bea
U_{n-1}&=&f_n^2+f_n^\prime+\omega^2_n\,,\label{2-3a}\\
U_{n}&=&f_n^2-f_n^\prime+\omega^2_n\,,\label{2-3b}
\eea
\eml
where $\omega_N=\omega_t=0$ and $U_N(x)=U(x)$. Once one knows
$U_{n-1}(x)$, the definition
\be
\label{2-4}
f_n(x)=\frac{w_n^\prime(x)}{w_n(x)},
\ee
can be used together with Eq.~(\ref{2-3a}) to obtain the second-order differential equation
\be
\label{2-5}
-\frac{d^2w_n}{dx^2}+\left[U_{n-1}(x)-\omega^2_n\right]w_n(x)=0\,.
\ee
The solutions $w_n(x)$ must be always unbounded in the limit
$x\to\pm\infty$. This is because (\ref{2-5}) is a Sch\"odinger-like
equation for the zero energy state, which must be absent in the
spectrum of $U_{n-1}-\omega_n^2$.

At this point, one notes that there is another freedom of choice
that resides beyond the knowledge of the complete discrete
spectrum: the choice of the potential $U_0(x)$, which is
arbitrary. The only condition over $U_0$ is to have a positive
nonzero continuum spectrum. This arbitrariness is eliminated if
one wants reflectionless symmetric potentials $U(x)$. In such a
case, $U_0$ must be a positive constant \cite{Bordag:2002pm,Vachaspati:2003vk}.

\section{Reconstruction from reflectionless potentials}
\label{section3}
The reconstruction of field theory from symmetric reflectionless
potentials has already been studied for the case of potentials that support
one bound state and two bound states \cite{Bordag:2002pm,Vachaspati:2003vk}. In this
section we shall revisit such problem without fixing, at first,
the constant $c$ in \eqref{2-1} and verify what happens in both
cases.

\subsection{One bound state}
\label{sinogordon}

The situation with one bound state has only the eigenvalue
$\omega^2_1=0$. Reflectionless symmetric potentials demand $U_0$
to be a positive constant potential. So, we may write
$U_0=\alpha^2>0$, for $\alpha$ real. One of the solutions of the differential
equation (\ref{2-5}) is
\be
\label{3-1}
w_1(x)=\cosh(\alpha x)\,.
\ee
The other solution is $\sinh(\alpha x)$, but it is antisymmetric and will not be considered here.
A suitable integration constant is not necessary, since it would not appear in (\ref{2-4}) nor in the potential
$U(x)$. Applying (\ref{3-1}) to both (\ref{2-4}) and (\ref{2-3b})
gives
\be
\label{3-2}
f_1(x)=\alpha\tanh(\alpha x),
\ee
and
\be
\label{3-3}
U(x)=\alpha^2\left[1-2\sech^2(\alpha x)\right]\,.
\ee
By taking the change of variable $\alpha x\to x$ in the associated
Schr\"odinger Eq.~(\ref{9}), the potential can be rewritten as
\be
\label{3-31}
U(x)=1-2\sech^2(x)\,.
\ee
The field $\phi$ is obtained from (\ref{2-1}) and reads
\be
\label{3-4}
\phi(x)=\pm
2\arctan\left[\tanh\left(\frac{x}{2}\right)\right]-c\,.
\ee
Then, by inverting \eqref{3-4} to obtain $x(\phi)$ in
(\ref{2-2}) one obtains
\be
\label{3-5}
V(\phi)=\frac{1}{2}\cos^2\left(\phi+c\right)\,.
\ee
Note that $V(\phi)$ has minima when $\phi+c=(2n+1)\pi/2$ for
integer $n$. The complete set of solution then reads
\be
\label{3-55}
\phi_n(x)=\pm
2\arctan\left[\arctanh\left(\frac{x}{2}\right)\right]+n\pi\,.
\ee
Here we have chosen $c=n\pi$. Any other choice of $c$ in
\eqref{3-5} will rely into equivalent field theory, with the same
number of topological sectors. With the choice \eqref{3-55} one
has
\be
V(\phi)=\frac{1}{2}\cos^2\left(\phi\right),
\ee
which is the Sine-Gordon model. So far no novelty arises due to the
constant $c$ except that its different values describe solutions in different
topological sectors of the same theory.

\subsection{Two bound states}

As was claimed in the abstract, we shall see new information
arising from the choice of $c$ for the situation with two bound states.
Here, the two energy eigenvalues obey $\omega_1^2>\omega^2_2=0$.
If one uses (\ref{2-5}) and defines $U_0-\omega_1^2=\alpha^2>0$,
it is possible to write
\be
\label{3-6}
w_1(x)=\cosh(\alpha x),
\ee
which gives
\be
\label{3-7}
f_1(x)=\alpha \tanh(\alpha x)\,.
\ee
As before, we set $U_0$ as a positive constant and dropped the
linearly independent solution $\sinh(\alpha x)$ to obtain
symmetric reflectionless potentials. The resulting potential
$U_1(x)$ is
\be
\label{3-8}
U_1(x)=\alpha^2\left[1-2\sech^2(\alpha x)\right]+\omega_1^2\,.
\ee
For $U_2(x)$ one uses \eqref{2-5} once again and solves the equation
\be
\label{3-9}
-\frac{d^2w_2}{dx^2}+\left[\alpha^2+\omega_1^2-2\alpha^2\sech^2(\alpha
x)\right]w_2=0\,.
\ee
By performing the change of variable $y=\tanh(\alpha x)$ one
arrives at the associated Legendre differential equation for
$l=1$~\cite{Zwillinger}, that is to say,
\be
\label{3-10}
\frac{d}{dy}\left[(1-y^2)\frac{d
}{dy}w_2\right]+\left(2-\frac{m^2}{1-y^2}\right)w_2=0\,,
\ee
where
\be
\label{3-11}
m^2\equiv 1+\frac{\omega^2_1}{\alpha^2}\,.
\ee
Since one needs $w_2(x)$ to be a nonlimited symmetric function,
the suitable solutions of (\ref{3-10}) are the
Legendre polynomials of the second kind, $Q_1^m(y)$, for $m=2,4,6,\cdots$. Here, we
need some extra information about the system in order to be able
to reconstruct $U(x)$ univocally. In other words, one has to know
the difference between $U_0$ and $\omega^2_1$, and for the
purpose of the current work, we choose $m=2$. In this case one gets
$\omega^2_1=3\alpha^2$,
\be
\label{3-12}
w_2(x)=\cosh^2(\alpha x)\,,
\ee
\be
\label{3-13}
f_2(x)=2\alpha\tanh(\alpha x)\,,
\ee
and
\be
\label{3-14}
U(x)=\alpha^2\left[4-6\sech^2(\alpha x)\right]\,,
\ee
since here $U_2(x)$ must be identified with $U(x)$. One can redefine $\alpha x\to x$\,
such that the Schr\"odinger equation (\ref{9}) associated with
(\ref{3-14}) will have the potential
\be
\label{3-142}
U(x)= 4-6\sech^2(x)\,,
\ee
eliminating $\alpha$ in Eqs. (\ref{3-12}) and (\ref{3-13}). As
expected, the translational mode will be a square integrable
function
\be
\label{3-15}
\eta_t(x)=N\sech^2\left(x\right)\,,
\ee
while
\be
\label{3-16}
\phi(x)=\pm \tanh\left(x\right)-c\,.
\ee
By inverting $\phi(x)$ in \eqref{3-16} and applying $x(\phi)$ to
(\ref{2-2}) one gets
\bea
\label{3-17}
V(\phi)&=&\frac{1}{2}\sech^4\left(x\right)\nonumber\\
&=&\frac{1}{2}\left[1-\phi^2-2c\phi-c^2\right]^2\,.
\eea

This result requires a closer investigation, to study the different field theories that may
emerge due to distinct choices of $c$. If one recalls the discussion leading to Eq.~\eqref{check}, here
we should probe the cases $c=0$, and also $c=\pm1$, as we consider below.

\subsubsection{$c=0$}

The choice $c=0$ reproduces the result of
\cite{Bordag:2002pm,Vachaspati:2003vk} and is the well-known
$\phi^4$ theory characterized by the potential
\be\label{3-20}
V(\phi)=\frac{1}{2}\left(1-\phi^2\right)^2\,.
\ee
This choice for $c$ reduces the problem to a problem with one topological
sector since the potential $V(\phi)$ has only two minima
($\bar{\phi}_{\pm}=\pm 1$) and one topological sector described by
the kink and antikink solutions
\be
\label{3-21}
\phi(x)=\pm\tanh\left(x\right).
\ee

This construction is univocal, in the sense that a simple redefinition of the field
$\phi$ in Eq.~(\ref{3-20}) will not bring another model, because it has no power to modify the number
of minima nor the topological structure of the model. As argued before, this is also valid for the Sine-Gordon model.

\subsubsection{$c=\pm1$}

\begin{figure}[t]
\includegraphics[width=7.4cm]{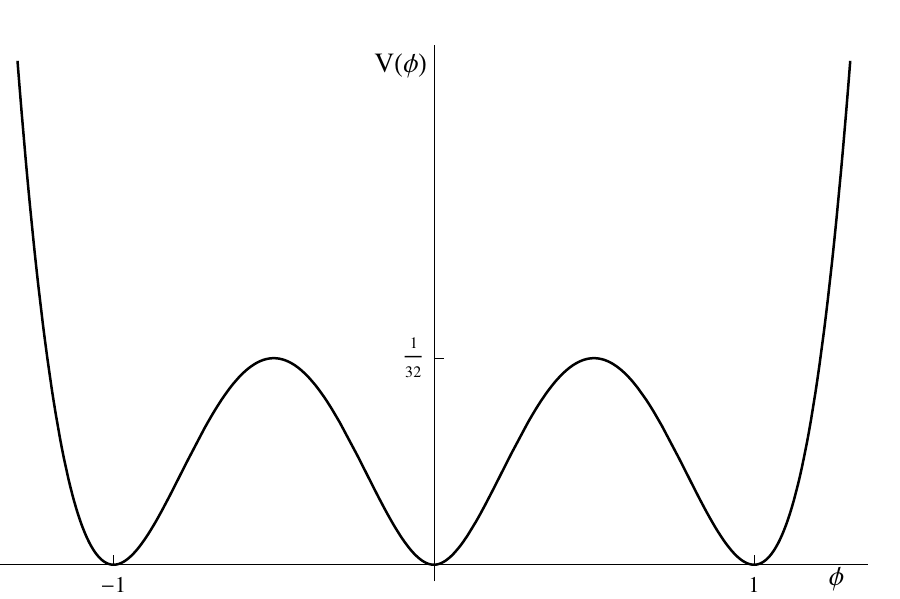}
\caption{The potential \eqref{3-23}, displayed to show how it behaves around its two local maxima, one
in each one of the two topological sectors.}\label{fig0}
\end{figure}

The choice $c=\pm1$ is different. It describes two distinct
topological sectors of the same field theory. Consider the
solutions
\be
\phi(x)=\pm\tanh(x)\mp1\,,
\ee
where the $\pm$ sign distinguishes between the two different
BPS solutions, while the $\mp1$ comes from the choice of $c$ as
$\pm1$ in Eq.~\eqref{3-16}. In this case, since $\phi=-|\phi|$ when $c=1$ and $\phi=|\phi|$ when
$c=-1$, one can easily verify that the resulting potential is the
same for both choices of $c$, namely,
\be
\label{3-22}
V(\phi)=\frac12{\phi^2}\left(2c+\phi\right)^2=\frac12{\phi^2}\left(2-|\phi|\right)^2\,.
\ee
We can redefine $\phi\to2\phi$ and $x^\mu\to x^\mu/2$ in the action \eqref{1} to get
the model with potential
\be
\label{3-23}
V(\phi)=\frac12{\phi^2}(1-|\phi|)^2\,.
\ee
It has three minima ($\bar{\phi}_0=0$ and $\bar{\phi}_{\pm}=\pm1$) and two
topological sectors, one connecting the minima $\bar{\phi}_-\leftrightarrow\bar{\phi}_0$ and
the other $\bar{\phi}_0\leftrightarrow\bar{\phi}_+$. This model was studied before
in \cite{Bazeia:2004ef}. It seems to mimic the $\phi^6$ theory studied in~\cite{Lohe1979},
but here the potential inside each one of its two topological sectors is symmetric around each
one of its local maxima. See Fig.~\ref{fig0}, where the potential is displayed.
This fact is important since it results in a symmetric reflectionless quantum
mechanical potential. Another fact of interest is that although this potential is of the fourth-order power in the field, its symmetric $(\bar\phi_0=0)$ and asymmetric $(\bar\phi_\pm=\pm1)$ minima suggest
the possibility of describing a first-order phase transition, and this cannot be described by the $\phi^4$ model
of Eq.~\eqref{3-20}.

In the $\phi^6$ case \cite{Lohe1979}, the potential is different and loses the symmetry behavior of the model \eqref{3-23},
leading to asymmetric potentials that are not reflectionless anymore. As a consequence, the $\phi^6$ model
cannot be obtained from the above reconstruction.

It is worth mentioning that a simple redefinition of the field
$\phi$ will not bring \eqref{3-23} to the
$\phi^4$ model. This happens because a shift in the field has no power
to modify the number of minima nor the topological structure of
the potential. This proves our claim that the reconstruction
procedure is not always univocal.

\section{Comments and conclusions}
\label{conclusions}
In this work we studied the reconstruction of field theory models from supersymmetric quantum mechanics. We addressed the case of potentials that support one and two bound states that are symmetric and reflectionless, as studied before in Refs.~\cite{Bordag:2002pm,Vachaspati:2003vk}.
We demonstrated that, although the problem with one bound state recovers univocally the sine-Gordon theory, the potential with two bound states does not result in a unique field theory. In this last case, we recovered two inequivalent field theories, the $\phi^4$ model that contains two minima and a single topological sector, and a modified $\phi^4$ model containing three minima and two topological sectors.

This work was motivated by the recent investigation \cite{Bazeia:2016xfc}, in which the authors studied the other route, the passage from field theory to supersymmetric quantum mechanics. There, it was shown that such a route is not unique, so we asked if this is true or not in the reconstruction process, in the passage from quantum mechanics to field theory. We explored this possibility in the current work, bringing an interesting new result which we think will motivate new investigations in the subject. We are now examining other systems, dealing with other potentials and possible generalizations. In particular,  we are studying the deformation procedure developed in \cite{Bazeia:2002xg}, to see how it can contribute to the reconstruction procedure. These and other related issues are currently under consideration, and we hope to report on them in the near future.

\begin{acknowledgments}
The authors would like to thank the Brazilian agencies CAPES and
CNPq for financial support. DB is also thankfull for support from the CNPq
grants 455931/2014-3 and 306614/2014-6.
\end{acknowledgments}


%

\end{document}